# Beam Dynamics Studies for a High Current Ion Injector[1]

A. Sauer, H. Deitinghoff, H. Klein, Institut für Angewandte Physik, Universität Frankfurt am Main, Germany

*Abstract*

Recent ion source developments resulted in the generation of high brilliance, high current beams of protons and light ions. After extraction and transport the beams with large internal space charge forces have to be captured, bunched and preaccelerated for the injection into the following driver part of a new generation of high intensity beam facilities for neutron sources, energy production, transmutation e.g. A combination of RFQ and DTL is considered to be a good solution for such a high current ion injector. Some preliminary beam dynamics layouts have been investigated by multiparticle simulations. Basic parameters like frequency, ion energy and sparking have been varied for the **I**nternational **F**usion **M**aterial **I**rradiation **F**acility (IFMIF) scenario as an example. The main interest was directed to high transmission, low losses and emittance conservation. The beam matching to the RFQ is shortly discussed as well as the matching between RFQ and DTL.

## 1 INTRODUCTION

During the last years the feasibility of a new generation of high intensity ion accelerator facilities has been studied. They consist of an rf linac, in some cases followed by rings for postacceleration, beam accumulation or pulse compression. Well known examples are drivers for spallation neutron sources, transmutation of radioactive waste, energy amplifier e.g. A smaller scale project is a material test machine (IFMIF) [1]. In all proposed schemes the linac starts with high current ion sources followed by RFQ accelerators and drift tube linacs as initial part, in high energy beam facilities coupled cavity linacs or synchrotrons take over for further acceleration. The requested beam peak currents are in the range of some 10 mA to more than 100 mA of light and heavy ions. A high quality of the beam formation in the injector part is essential for the whole complex, because small emittances and well confined bunches are needed to avoid losses and activation along the accelerator. One critical aspect is the matching of the beam between the different parts of the driver, which will be discussed in the following for IFMIF.

## 2 THE IFMIF INJECTOR PART

Fig. 1 gives a scheme of the proposed IFMIF facilities: Currents of up to 0.5 A D$^+$ ions with variable final energies between 16 and 20 MeV/u are used to produce very high fluxes of neutrons for testing the wall materials for magnetic fusion reactors.

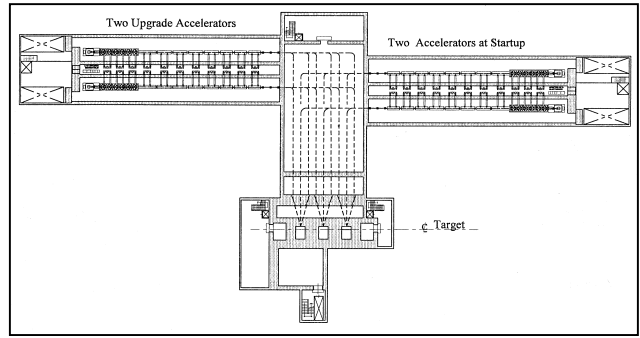

Figure 1: Basic layout of the IFMIF facility.

While in high energy machines the required high beam current at the injector end is generated by combining beams from 2 or more ion sources in funnelling steps, in the IFMIF case 2 to 4 accelerators in parallel are planned due to its comparatively small size. Each of them has to capture, focus and accelerate a beam of 125 mA in cw operation. Beam losses must be kept as small as possible to avoid activation, which becomes serious for beam energies higher than 1.0 MeV/u. A base line design has been worked out for the accelerator, parameters are listed in Table 1 [1].

Table 1: Base line parameters of the IFMIF injector

|  | Source | RFQ | DTL |
|---|---|---|---|
| $I_{out}$ [mA] | 150 | 140 | 125 |
| $W_{out}$ [MeV/u] | 0.05 | 4.0 | 20.0 |
| $\varepsilon^{RMS,N}_{out}$ [cm×mrad] | 0.02 | 0.04 | 0.04 |
| Length [m] | 4.2 | 11.7 | 30.4 |
| Frequency [MHz] | 175 | 175 | 175 |
| Beam Power [MW] | 5.0 each | | |

A first beam dynamics layout of the RFQ employing equipartitioning rules has been carried out successfully already in 1996 [2]. Also a design of a DTL for 4 MeV/u was made. Recently two parameters of this base line design have been reconsidered:

1) The electrode voltage assumed for the RFQ was rather high: applying the Kilpatrick criterion for rf breakdown in the form ,

$$f[MHz] = 1.643 \times 10^4 \cdot E_K^2 [MV/cm] \cdot e^{-\frac{0.085}{E_K}}, \quad [1]$$

---

[1] Work supported by the European Commission.

a Kilpatrick factor of 2.1 came out, which is reasonable for pulsed operation but considered as too high for cw operation of a structure. Therefore new designs have been attempted with a lower value of 1.7 Kilpatrick.

2) The RFQ is highly efficient in capturing and focusing high intensity beams, but the acceleration rate is rather low. Therefore the end energy of the RFQ of 4 MeV/u should be lowered to 2.5 MeV/u.

## 3 BEAM DYNAMICS LAYOUT OF RFQ AND DTL AT 175 MHZ

In the baseline design the RFQ has to accelerate a 125 mA Deuteron beam from 50 keV/u to 4 MeV/u. One possible layout was found at 2.1 Kilpatrick introducing some equipartitioning rules into the design. The calculated beam behaviour was very satisfying: high transmission, small emittance growth and a well compressed bunch of small phase width, which is well prepared for direct injection into the DTL.

Table 2: Design parameters of the different IFMIF 175 MHz RFQs

|  | 175 MHz RFQ design studies | | |
|---|---|---|---|
| **Ion** | $D^+$ | $D^+$ | $D^+$ |
| **Design** | EP (Deshan) | EP (Jameson) | Conserv. (Sauer) |
| **N** | 5000 | 5000 | 5000 |
| **Input dist.** | 4d Wbag | 4d Wbag | 4d Wbag |
| **F [MHz]** | 175 | 175 | 175 |
| **$W_{in}/W_{out}$ [MeV/u]** | 0.05 / 2.5 | 0.05 / 2.5 | 0.125 / 2.5 |
| **V [MV]** | 0.100 - 0.164 | 0.111 - 0.151 | 0.126 |
| **b** | 2.13 | 1.7 | 1.7 |
| **$RMS_{in}$** | 4 | 4 | 6 |
| **$RMS_{out}$** | Crandall cell | Crandall cell | Crandall cell |
| **C / L [m]** | 382 / 8.21 | 660 / 12.32 | 565 / 14.52 |
| **$\phi_{syn}$ [°]** | -90. - -31. | -90. - -35.3 | -90. - -37. |
| **$I_{in}$ [mA]** | 140 | 140 | 140 |
| **$I_{out}$ [mA]** | 133.2 | 129.7 | 127.50 |
| **Tr** | 95,2% | 92,7% | 91,0% |
| **m** | 1.00 – 1.78 | 1.00 – 1.74 | 1.00 – 1.62 |
| **$\varepsilon^{N,RMS}_{trans}$ [cm×mrad]** | 0.020 / 0.030 | 0.020 / 0.0226 | 0.021 / 0.0255 |
| **$\varepsilon^{N,RMS}_{long}$ [cm×mrad]** | 0 / 0.057 | 0 / 0.046 | 0 / 0.054 |

Lowering the electrode voltage corresponding to a Kilpatrick factor of 1.7 made new designs necessary, at the same time the output energy of the RFQ was fixed to 2.5 MeV/u. Up to now the most promising design was found by R. Jameson employing equipartitioning again [3]. The third solution of Table 2 (Sauer) with simpler design methods following the rules of CURLI and RFQUICK have been obtained at higher input energies, nevertheless showing, that an RFQ design with a constant and modest voltage level is feasible. All examples from Table 2 show low losses and low emittance growth but a larger phase width of the bunch, the same is true for the 2.1 Kilpatrick design, if cut at 2.5 MeV/u. The matching of the RFQ input emittances has been done with the help of PTEQ [4] which delivers optimised Twiss parameters of the ellipses and a smoothed beam behaviour along the RFQ. This is an idealized assumption: Former investigations on the matching to an RFQ with the help of a magnetic LEBT have shown [5], that small changes in current, solenoid fields and emittances in the order of a view percent do not cause losses but generate a large emittance growth in the RFQ. Therefore quite larger emittances than the ideal ones may be expected at the RFQ end.

A provisional layout of an DTL has been done with SUPERFISH and PARMILA, to check the matching possibilities at the lower input energy of 2.5 MeV/u. The design was made for constant transverse focusing, constant rf field amplitude and some initial phase ramping. In the very first part of the DTL the filling factor had to be risen to 0.59 at pole tip fields below 1.0 T. Afterwards it was kept to 0.5 and lower. After some initial checks of the beam dynamics with matched input emittances generated by the program TRACE3D which showed a good beam behaviour, RFQ output emittances calculated with PARMTEQM were taken as an input to PARMILA. In all cases a short drift of $\beta\lambda/2$ (6.2 cm) was assumed between RFQ and DTL. Emittance growth was unavoidable in that case. Fig. 2 shows the calculated output emittances of the DTL at 20 MeV/u, with RFQ 3 of Table 2 as the previous accelerator.

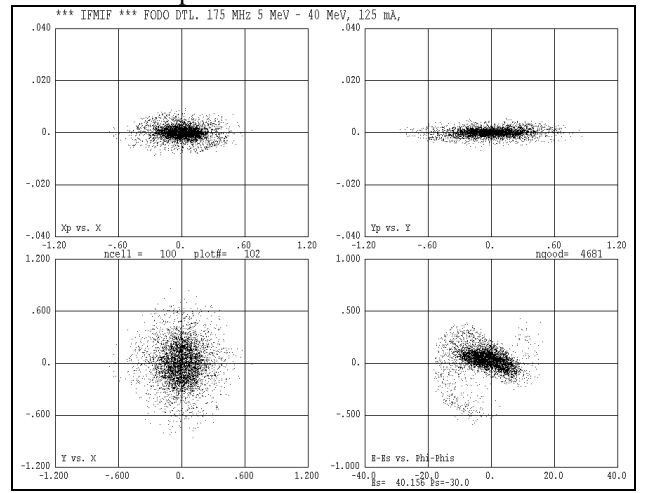

Figure 2: Output emittance of the 175 MHz DTL with direct input from the RFQ 3 from Table 2 and a 6.2 cm drift between RFQ and DTL.

## 4 MEBT DESIGN

In the following a compact matching section (MEBT) with a length below 1.0 m is discussed. The design was made with TRACE3D which calculates the optimised field strength of the four magnetic quadrupoles (for transverse matching) and the two rebuncher cavities (for longitudinal matching). The result is a mismatch factor between RFQ and DTL smaller than 1 from first 2.5 without MEBT. Transfer of this MEBT to PARMILA and injection of the RFQ output emittances via matching line into the DTL showed a large improvement in the beam behaviour as can be seen in comparing Fig. 2 and Fig. 3.

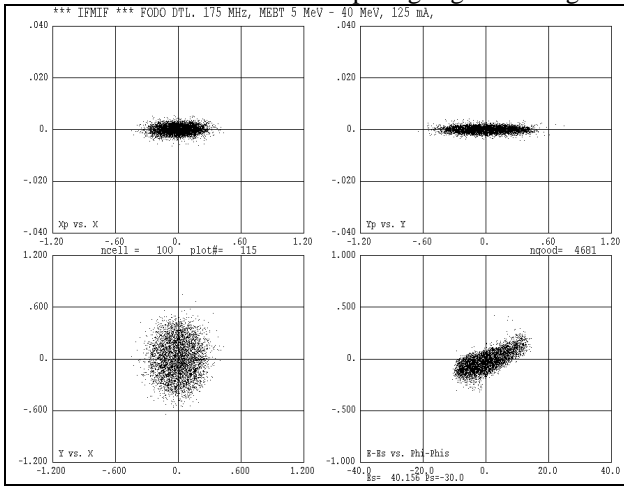

Figure 3: Output emittances of the 175 MHz DTL with a MEBT of 0.686 m between RFQ and DTL.

The beam is quite nicely matched along the DTL with an rms emittance growth of 10 %, no losses and well shaped output emittances. Moreover such a design allows the installation of steerers as well as beam diagnostics and of a valve and bellows, which is important in the practical view of an accelerator facility [7].

## 5 BEAM DYNAMICS LAYOUT OF RFQ AND DTL AT 120 MHZ

Layout of RFQ and DTL have been repeated for a lower frequency of 120 MHz, which gives for the 4-Rod-RFQ still small geometrical dimensions [6]. The advantage is a lower rf defocusing term and higher focusing efficiency, therefore RFQ layouts for 1.7 Kilpatrick are more easily reached for the reference energy also with the conventional design method. A transmission of more than 90 % with 40 % rms emittance growth were achieved. At 2.5 MeV/u the relation between longitudinal output emittance of the RFQ and the acceptance of the DTL is nearly the same as in the case at 175 MHz. Again the length of the RFQ output bunch of about 70° is too large for direct injection into the DTL. With the same matching procedure as before using TRACE3D and $\beta\lambda/2$ (which is now 9.1 cm) as basic length for drift, rebuncher cavities and quads a MEBT of 1.0 m has been designed for a low mismatch factor smaller than 1 again. Fig 5. shows the output emittances after the DTL, when the calculated output emittances of the 120 MHz RFQ were taken as input to MEBT and DTL for PARMILA. Again 100% transmission and an rms emittances growth of about 10% only were obtained for MEBT and DTL.

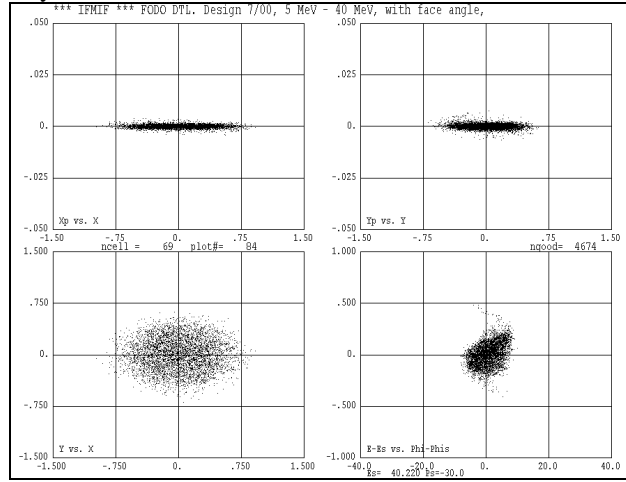

Figure 4: Output emittances of the 120 MHz DTL with a MEBT of 1.001 m between RFQ and DTL.

## 6 CONCLUSIONS

In the IFMIF scenario a beam dynamics layout for 125 mA $D^+$ RFQ at 1.7 Kilpatrick. was successfully carried out for both frequencies 175 MHz and 120 MHz resp. While for a RFQ with high end energy and a very well bunched output beam with a small phase width of about ±20° a direct injection into the DTL shows smooth beam behaviour, low emittance growth and no losses. A matching line seems to be favourable in the case of a shorter RFQ with lower end energy and larger phase width of the output bucket to prevent filamentation and large emittance growth in the beam. In addition a matching line provides the possibility for beam diagnostics and adjusting devices. The advantages and disadvantages of both cases have to be studied carefully, especially the sensitivity against errors and initial mismatch, which may strongly influence the size and shape of the RFQ output beam.

## REFERENCES

[1] IFMIF Final Report, December 1996 (1997).
[2] D. Li, R.A. Jameson, "Particle Dynamics Design Aspects for an IFMIF $D^+$ RFQ", EPAC'96, Sitges, June 1996.
[3] R.A. Jameson, "Some characteristics of the IFMIF RFQ KP1.7 designs", IFMIF Memo, RAJ-24-May-2000.
[4] R.A. Jameson, "A discussion of RFQ LINAC Simulation", LA-CP-97-54, September 1997.
[5] A. Sauer, Dipl. Thesis, Universität Frankfurt, 1998.
[6] A. Schempp, "Advances of Accelerator Physics and Technologies", Ed. H. Schopper, World Sci., 1993.
[7] U. Ratzinger, private communications, August 2000.